# Jordan Journal of Physics

## REVIEW ARTICLE

## Fowler-Nordheim Plot Analysis: a Progress Report

Richard G. Forbes[a], Jonathan H. B. Deane[b], Andreas Fischer[c] and Marwan S. Mousa[d]

[a] *Advanced Technology Institute & Department of Electronic Engineering, Faculty of Engineering and Physical Sciences, University of Surrey, Guildford, Surrey GU2 7XH, UK.*

[b] *Department of Mathematics, Faculty of Engineering and Physical Sciences, University of Surrey, Guildford, Surrey GU2 7XH, UK.*

[c] *Institut für Physik, Technische Universität Chemnitz, Germany.*

[d] *Department of Physics, Mu'tah University, Al-Karak 61710, Jordan.*

**Abstract:** The commonest method of characterizing a cold field electron emitter is to measure its current-voltage characteristics, and the commonest method of analysing these characteristics is by means of a Fowler-Nordheim (FN) plot. This tutorial/review-type paper outlines a more systematic method of setting out the Fowler-Nordheim-type theory of cold field electron emission, and brings together and summarizes the current state of work by the authors on developing the theory and methodology of FN plot analysis. This has turned out to be far more complicated than originally expected. Emphasis is placed in this paper on: (a) the interpretation of FN-plot slopes, which is currently both easier and of more experimental interest than the analysis of FN-plot intercepts; and (b) preliminary explorations into developing methodology for interpreting current-voltage characteristics when there is series resistance in the conduction path from the high-voltage generator to the emitter's emitting regions. This work reinforces our view that FN-plot analysis is best carried out on the raw measured current-voltage data, without pre-conversion into another data format, particularly if series resistance is present in the measuring circuit. Relevant formulae are given for extracting field-enhancement-factor values from such an analysis.

## 1. General Introduction

*Field electron emission (FE)* is the field induced emission of electrons from a solid or liquid emitter. Field electron sources have a number of actual or potential applications, including (in their single-tip form) small bright sources for electron microscopes and similar electron beam instruments, and (in their large-area form) extended sources for electronic devices, microwave generators or spacecraft neutralizers. FE is also a potential primary cause of electrical breakdown in vacuum, and needs to be understood so that breakdown can be avoided. There is a technological need for reliable characterization of field emitters.

The commonest method of investigating emitter behaviour is to measure current-voltage ($i_m$-$V_m$) characteristics and extract one or more characterization parameters from a Fowler-Nordheim (FN) plot (i.e., a plot of the form $\ln\{i_m/V_m^2\}$ vs $1/V_m$, or equivalent using other variables). In recent years, the authors of this paper have been developing theory aimed at improving the methodology of FN-plot analysis. In detail, this has turned out to be a very intricate and algebraically complicated problem, far more so than originally expected. It has involved repeated improvement and reformulation of theoretical approaches, as details of the overall problem become clearer.

The present paper is, in essence, a tutorial/progress report relating to FN plot analysis. It brings together in a single place and summarises our previous work on this topic. It aims to set out relevant theory as we currently understand and formulate it, and to provide a brief report on progress made towards more complete understanding. This report focuses more on FN-plot slope values than on intercept values. This is partly because slope data have to be properly understood before reliable interpretations of intercept data can be given, partly because slope data are currently of greater experimental interest. For both reasons, understanding how to analyze slope data is more advanced than understanding intercept data.

The main way in which this account differs from earlier treatments is that more attention is given to series-resistance effects in the measurement circuit. These are a common cause of "saturation" effects in FN plots, and are probably the commonest cause of unreliable results.

More generally, this paper should be regarded as updating and (to some extent) replacing earlier discussions of FN-plot theory given by the present authors. Its structure is as follows. Section 2 sets out background theory. Section 3 discusses current-voltage data-analysis using FN plots. Section 4 begins to explore how to analyze current/voltage data gathered from circuits that contain significant series resistance. Section 5 indicates work that remains to be done.

The normal electron emission convention is employed that fields, voltages, currents and current densities are treated as positive, even where they would be regarded as negative in classical electromagnetism.

## 2. Theoretical Background
### 2.1 Underlying assumptions

Basic tunnelling theory is formulated in the ideal theoretical context of the so-called *bulk emitter*, which fills half of space and has a smooth flat planar surface, with an uniform external electrostatic field $F_{ext}$ in vacuum above the emitter. Atomic structure is ignored and a Sommerfeld-type free-electron model [1] is assumed for the emitter, with the electron population taken to obey Fermi-Dirac statistics and to be in thermodynamic equilibrium at thermodynamic temperature $T$. In the simplest models $T$ is taken as 0 K. At low to moderate temperatures, emission is only very weakly dependent on temperature; thus, zero-K models are adequately applicable up to well above room temperature.

It is then assumed that, provided the local radius of curvature is not too small (greater than about 10 to 20 nm), and if $F_{ext}$ is replaced by the *local barrier field* $F_L$, where $F_L$ is the electrostatic field at point "L" in the emitter's electrical surface [2], then bulk-emitter theory is applicable to the electron emission near point L. This field $F_L$ helps specify the local tunnelling barrier for electrons that "escape from point L".

In practice, with real emitters, interest is usually in the *characteristic (local) emission current density (ECD) $J_C$* at some point "C" in the electrical surface chosen to be characteristic of the emitter. For *single-tip field emitters (STFEs)* and models thereof, it is often convenient to take C at the emitter apex; for *large-area field emitters (LAFEs)* point C can be thought of as the point where the local ECD is highest for a given voltage, which will often be at the apex of one of the individual emitting sites. Parameters subscripted "C" relate to point C.

## 2.2 Basic tunnelling theory

*Fowler-Nordheim (FN) tunnelling* is wave-mechanical tunnelling through an exact or rounded triangular barrier. *Deep tunnelling* is tunnelling at a forwards energy level well below the top of the barrier, at a level where the Landau & Lifshitz approximation [3,4] of general tunnelling/ transmission theory is adequately valid. *Cold field electron emission (CFE)* is a statistical emission regime where most electrons escape by deep tunnelling from electron states close to the emitter Fermi level.

For electron motion along a coordinate $z$, in a situation where the single-particle three-dimensional time-independent Schrödinger equation separates in Cartesian coordinates, the equation for the wave-function component $\psi_z(z)$ can be written

$$\left.\begin{aligned} d^2\psi_z/dz^2 &= -(2m/\hbar^2)[U(z)-E_z]\,\psi_z(z) \\ &\equiv -\kappa^2 M(z)\,\psi_z(z) \end{aligned}\right\}, \quad (1)$$

where $\hbar$ is Planck's constant divided by $2\pi$, $m$ is the electron mass in free space, $\kappa$ [$\equiv (2m)^{1/2}/\hbar$] is a universal constant defined by FN [5,6], $U(z)$ is the total electron potential energy, and $E_z$ is the total-electron-energy component associated with motion in the $z$-direction. The quantity $M(z)$ is defined as $[U(z)-E_z]$, and is termed the *electron motive energy*.

A one-dimensional *wave-mechanical barrier* is a region of space (along the $z$-coordinate) where $M(z)$ is continuously positive (where the kinetic energy of a hypothetical classical point particle would be continuously negative). This barrier is characterized by a parameter $G$ defined by

$$G = 2\kappa \int M^{1/2}(z)\,dz \equiv g_e \int M^{1/2}(z)\,dz, \quad (2)$$

where the integral is taken "across the barrier" (i.e., across the region where $M(z) \geq 0$), and $g_e$ [$\equiv 2\kappa = (8m)^{1/2}/\hbar$] is a universal constant [6]. The parameter $G$ has been called both the "Gamow exponent" and the "JWKB exponent", but the physical name *barrier strength* is now preferred. (Strong barriers are difficult to tunnel through.) Eq. (2) can be called the *barrier-strength integral*.

In the *Landau and Lifschitz approximation* [3, 4], the *tunnelling probability D* that an electron approaching the barrier escapes through it is given by

$$D \approx P \exp[-G], \quad (3)$$

where $P$ is a *tunnelling pre-factor*. Except in special cases, $P$ is very difficult to calculate accurately, but is thought to typically have values in the range 0.4 to 1.1 (see Appendix A). In the related *simple-JWKB approximation*, which is the approximation normally used, $P$ is set equal to unity, and Eq. (3) becomes

$$D \approx \exp[-G]. \quad (4)$$

## 2.3 Barrier form and related topics

Physically, an abstract expression for the motive energy $M(z)$ is

$$M(z) = H + U^{ES}(z) + U^{XC}(z), \quad (5)$$

where $H$ is a constant called the *zero-field barrier height*, and $U^{ES}$ and $U^{XC}$ are terms associated, respectively, with electrostatic (ES) and with exchange-and-correlation (XC) effects. The detailed mathematical form of $M(z)$ defines the *barrier form* (or "shape").

The simplest barrier exhibiting FN tunnelling is the *exactly triangular (ET) barrier* defined by disregarding $U^{XC}$, taking $U^{ES} = -eF_L z$, and writing $M^{ET}(z) = H - eF_L z$. Here, $e$ is the elementary positive charge, $z$ is distance measured from the emitter's electrical surface, and $F_L$ is the local barrier field, as defined above. For this barrier, the barrier strength $G^{ET}$ is easily shown to be

$$G^{ET} = bH^{3/2}/F_L, \quad (6)$$

where $b$ [$\equiv 2g_e/3e = (4/3)(2m)^{1/2}/e\hbar$] is the *second Fowler-Nordheim constant* [6].

For a "general barrier" (GB) of the same zero-field height $H$ but otherwise of arbitrary but well behaved form, the barrier strength $G^{GB}$ can be evaluated from Eq. (2), by numerical integration if necessary. A *barrier-form correction factor* $\nu^{GB}$ ("nu$^{GB}$") can then be defined via

$$G^{GB} = \nu^{GB} G^{ET}. \tag{7}$$

The simplest barrier including exchange-and-correlation effects is the *Schottky-Nordheim (SN) barrier*, which models these XC effects by *Schottky's planar image potential energy* $U^{XC} = -e^2/16\pi\varepsilon_0 z$ [7], where $\varepsilon_0$ is the electric constant. Adding this term to the ET barrier gives the motive energy for the SN barrier, namely

$$M^{SN}(z) = H - eF_L z - e^2/16\pi\varepsilon_0 z. \tag{8}$$

The maximum value, $M^{SN}(\max)$, of this function defines the *(reduced) barrier height*. It is readily shown that

$$\begin{aligned} M^{SN}(\max) &= H - (e^3/4\pi\varepsilon_0)^{1/2} F_L^{1/2} \\ &\equiv H - c_S F_L^{1/2} \end{aligned}, \tag{9}$$

where $c_S [\equiv (e^3/4\pi\varepsilon_0)^{1/2}]$ is a universal constant sometimes called the *Schottky constant* [6]. Clearly, for fields larger than the field at which $M^{SN}(\max)$ becomes zero, the tunnelling barrier vanishes.

Barriers with $H$ equal to the *local work function* $\phi$ play a special role in CFE theory. Eq. (9) shows that, for a barrier of zero-field height $\phi$, the *reference field* $F_R$ needed to make $M^{SN}(\max)$ zero is $F_R = c_S^{-2}\phi^2$. For the SN barrier, a parameter $f$ called the *scaled barrier field* (for a SN barrier of zero-field height $\phi$) can then be defined by

$$\begin{aligned} f &\equiv F_L/F_R = c_S^2 \phi^{-2} F_L = e^3 F_L/4\pi\varepsilon_0 \phi^2 \\ &\approx 1.439964 \, (\text{eV}/\phi)^2 \, (F_L/\text{V nm}^{-1}) \end{aligned}. \tag{10}$$

Clearly, the barrier height becomes zero when $f=1$. This relatively recently introduced parameter $f$ plays an important role in modern CFE theory, as indicated below.

## 2.4 Fowler-Nordheim-type equations

*Fowler-Nordheim-type (FN-type) equations* are a large family of approximate equations originally derived to describe CFE from bulk metals. As discussed below, FN-type equations can be formulated using many different sets of independent and dependent variables, and using many different detailed approximations. The *core theoretical formulation* gives the local ECD $J_L$ in terms of the local work function $\phi$ and the local barrier field $F_L$. This core formulation is obtained by summing ECD contributions from all relevant emitter electron states, and writing the result in the form $J_L = Z_F D_F$, where $D_F$ is the tunnelling probability for a barrier of zero-field height $\phi$, and $Z_F$ represents the related *effective electron supply* (effective incident current density). Here and elsewhere, the subscript label "F" indicates that a parameter relates to a Fermi-level electron moving "forwards" (i.e., towards and normal to the emitter surface) and/or to the barrier of zero-field height $\phi$ seen by this electron.

When written out explicitly, this core result is usefully put in the form of the linked equations

$$J_L = \lambda_L^{GB} J_{kL}^{GB}, \tag{11a}$$

$$J_{kL}^{GB} \equiv a\phi^{-1} F_L^2 \exp[-\nu_F^{GB} b\phi^{3/2}/F_L], \tag{11b}$$

where $a \, [\equiv e^3/8\pi h_P]$ is the *first FN constant* [6]. $h_P$ is Planck's constant. $J_{kL}^{GB}$ is the *local kernel current density* (for the general tunnelling barrier), and is a mathematical quantity defined by Eq. (11b).

The *local pre-exponential correction factor* $\lambda_L^{GB}$ is a correction factor that (in principle) takes account of all other relevant physical effects that influence the emission process, including: the tunnelling pre-factor $P$, exact integration over emitter states, temperature, use of atomic wave-functions, and band-structure effects. A (quantitatively unknown) factor that allows for failure to use the correct barrier form, and/or for any other unrecognized mathematical or physical inadequacy in the assumed theory, can also be included as contributing to $\lambda_L^{GB}$. In any particular CFE model, the predicted form and value of $\lambda_L^{GB}$ depend on precisely what theoretical assumptions are made. Exact realistic prediction of $\lambda_L^{GB}$-values would be intensively difficult, is in many cases beyond the present capabilities of quantum mechanics, is of limited economic value, and is unlikely to happen in the foreseeable future.

The merit of splitting Eq. (11) as above is that, given particular choices of the barrier form and of the values of $\phi$ and $F_L$, the quantity $J_{kL}^{GB}$ can be calculated *exactly*; thus, all theoretical uncertainties are accumulated into the parameter $\lambda_L^{GB}$. For metal emitters, assuming tunnelling through a SN barrier, our current best guess (in 2015) is that $\lambda_L^{GB}$ lies in the range $0.005 < \lambda_L^{GB} < 11$ (see Appendix A), but this could be an underestimate of the uncertainty range.

As indicated above, these equations apply adequately to metal STFEs, provided that the emitter is "not too small and sharp" (apex radius greater than about 10 to 20 nm). The exact limits of applicability of FN-type equations have never been definitively established.

With the same constraint, FN-type equations can also be used to analyze CFE data from LAFEs consisting of many individual emitters and/or emission sites. FN-type equations are also widely used to analyze experimental data relating to non-metallic materials. In any particular non-metallic case, the validity of doing this is open to question; currently, for non-metals, there are no reliable systematics as to when such an approach is adequate.

## 2.5 Complexity levels

In developing FN theory, many different detailed assumptions and approximations can be (and have been) made about the physical origins of and the mathematical forms of the correction factors $v_F^{GB}$ and $\lambda_L^{GB}$ in Eq. (11). The assumptions made determine the *complexity level* of the resulting FN-type equation. For emitters that are "not too small and sharp", the main complexity levels used historically and recently are shown in Table 1.

The name "*new-standard*" is introduced here, to cover equations that are based on the SN barrier but have a general form for the pre-exponential correction factor. A special case of this is the *orthodox* FN-type equation, where the additional mathematical assumption is made that, apart from the independent variable itself, the only parameter in the equation that varies significantly with the independent variable is the barrier form correction factor.

Note that different choices as to barrier form would lead to different deductions, from experiment, about the value of the pre-exponential correction factor. Hence this correction factor is formally different for each different choice of barrier form.

TABLE 1. Complexity levels of Fowler-Nordheim-type equations.

| Name | Date | Ref. | $\lambda_C^{GB} \rightarrow$ | Barrier form | $v_F^{GB} \rightarrow$ | Note |
|---|---|---|---|---|---|---|
| Elementary | ? | ? | 1 | ET | 1 | a |
| Original | 1928 | [5] | $P_{FN}$ | ET | 1 | b |
| Fowler-1936 | 1936 | [9] | 4 | ET | 1 | |
| Extended elementary | 2015 | here | $\lambda_C^{ET}$ | ET | 1 | |
| Dyke-Dolan | 1956 | [10] | 1 | SN | $v_F$ | c |
| Murphy-Good | 1956 | [12] | $t_F^{-2}$ | SN | $v_F$ | c |
| Orthodox | 2013 | [13] | $\lambda_C^{SN0}$ | SN | $v_F$ | d |
| New-standard | 2015 | here | $\lambda_C^{SN}$ | SN | $v_F$ | - |
| "Barrier-effects-only" | 2013 | [14] | $\lambda_C^{GB0}$ | GB | $v_F^{GB}$ | d |
| General | 1999 | [8] | $\lambda_C^{GB}$ | GB | $v_F^{GB}$ | |

[a]Many earlier imprecise versions exist, but the first clear statement seems to be in 1999 [8].
[b]For details concerning the Fowler-Nordheim tunnelling pre-factor $P_{FN}$, see [5] and [6].
[c]For modern theory concerning $v_F$ and $t_F^{-2}$, see [11].
[d]The superscript " $^0$ " indicates that the factor is to be treated mathematically as constant.

The name of the complexity level applies to the related equation for the characteristic local ECD $J_C$, as in Table 1, and also applies to the equivalent equations for the emission current $i_e$, the measured current $i_m$, and macroscopic current density $J_M$.

## 2.6 Scaled form for the kernel current density for the SN barrier

In the case of the SN barrier, the barrier form correction factor $v_F^{SN}$ is given by the particular value $v(f)$ ["vee(*f*)"] of a special mathematical function $v(l')$ called *the principal SN barrier function*, where $l'$ is a mathematical variable [15]. For mathematical convenience, this particular value $v(f)$ is sometimes denoted by $v_F$, but the two symbols mean the same thing. Exact and approximate expressions for $v(f)$ are known [11,15]. The simple approximation

$$v_F \equiv v(f) \approx 1 - f + (f/6)\ln f \qquad (12)$$

is valid to better than 0.33% over the whole range $0 \leq f \leq 1$, and is adequate for most technological purposes.

Older approximate formulae for $v_F$, and related evaluations, exist in the literature (e.g., [16]), but these are often given in terms of the Nordheim parameter $y$ [$=f^{1/2}$], and the approximate formulae are usually less accurate than Eq. (12). There are good physical and mathematical reasons [11] for the modern practice of using $f$ (or $l'$), rather than $y$; one good reason is the *linearity* of the relationship between $f$ and $F_L$, as given by Eq. (10).

One may define work-function-related parameters $\eta(\phi)$ and $\theta(\phi)$ by

$$\left.\begin{array}{r}\eta(\phi) \equiv b\phi^{3/2}/F_R = bc_S^2\phi^{-1/2} \\ \cong 9.836238\ (\mathrm{eV}/\phi)^{1/2}\end{array}\right\}, \qquad (13)$$

$$\left.\begin{array}{r}\theta(\phi) \equiv a\phi^{-1}F_R^2 = ac_S^{-4}\phi^3 \\ \cong (7.433979\times 10^{11}\ \mathrm{A/m^2})\,(\phi/\mathrm{eV})^3\end{array}\right\}. \qquad (14)$$

The local kernel current density $J_{kL}^{SN}$ for the SN barrier can then be written exactly *in scaled form* as

$$J_{kL}^{SN} = \theta f_L^2 \cdot \exp[-v(f_L)\cdot\eta/f_L]. \qquad (15)$$

Values of $\eta(\phi)$ and $\theta(\phi)$ are shown in table 2 in Ref. [13] for a range of work-function values. For illustration, when $\phi = 4.50$ eV, then $\eta \approx 4.637$ and $\theta \approx 6.77\times 10^{13}$ A/m². Merits of form (15) are that: (a) for a given $\phi$-value, only a single independent variable ($f_L$) appears in the right-hand side; and (b) a good simple approximation for $v(f_L)$ is known. Hence, good approximate values for $J_{kL}^{SN}$ are easy to obtain.

## 2.7 The universal FN-type equation

All FN-type equations can be seen as variants of the *universal FN-type equation*

$$Y = C_{YX} X^2 \exp[-B_X/X], \qquad (16)$$

where $X$ and $Y$ are universal independent and dependent variables, respectively, and represent any of the specific variables shown in Tables 2 and 3. By definition, a *FN-type equation* is an equation with the mathematical form of Eq. (16). The *form* of a FN-type equation (and of the corresponding FN plot) is specified by the particular choices of $X$ and $Y$. The theory of CFE as described by FN-type equations is sometimes called *Fowler-Nordheim theory*.

$C_{YX}$ and $B_X$ are parameters whose precise forms depend on the equation form and complexity level, and sometimes on other factors. $C_{YX}$ and $B_X$ may be weakly-to-moderately varying functions of $X$, and this variation will in some cases be significant. If the particular forms used for both $B_X$ and $C_{YX}$ are sufficiently general to encompass all physical effects associated with the particular choices of $X$ and $Y$ used, then the resulting equation is said to be *technically complete*.

Over the last few years it has become increasingly clear that there is a need to distinguish between (a) *emission variables* (i.e., the *emission current and voltage* $\{i_e, V_e\}$, and physical variables derived from them that relate directly to the geometry and electrostatics of the emitting device), and (b) *measured variables* {i.e., the *measured* current and voltage $\{i_m, V_m\}$, and mathematical variables derived algebraically from them). This leads to the three-way classification of variables shown in Table 2. To deal with the resulting complications, it is easiest to first set out the theory for the "theoretical" and "emission" variables.

At the orthodox complexity level, all independent theoretical and emission variables are linearly related to each other, and all dependent theoretical and emission variables are linearly related to each other. In this case, the parameter $f$ can be seen as a scaled form of any of the independent theoretical and emission variables, defined in any particular case by

$$f = X/X_R. \qquad (17)$$

where $X_R$ is the value of $X$ needed to reduce a SN barrier of zero-field height $\phi$ to zero.

## 2.8 Auxiliary parameters – independent variables

For the characteristic point "C", the exponent in Eq. (16) can be expanded in the forms

$$\left.\begin{array}{rl} B_X/X &= v_F^{GB} b\phi^{3/2}/F_C \\ &\equiv v_F^{GB} b\phi^{3/2}/c_X X \equiv v_F^{GB} B_X^{el}/X \end{array}\right\}, \quad (18)$$

where $B_X^{el} \equiv b\phi^{3/2}/c_X$. Parameters that interrelate different independent variables (or different dependent variables) are termed *auxiliary parameters*, and are defined via *auxiliary equations*. For the independent variables, Table 2 shows all auxiliary parameters and equations currently thought relevant, even though we would discourage the use of some of them. An important subset consists of those parameters and equations that relate the characteristic barrier field $F_C$ to the chosen independent variable: these have the general form

$$F_C = c_X X. \qquad (19)$$

Particular instances of $c_X$ are included in Table 2.

A specific issue is how best to write the auxiliary equation linking $F_C$ to the emission voltage $V_e$, because two different parameters ($\beta_{V,C}$ and $\zeta_C$) are available, as defined by

$$F_C \equiv \beta_{V,C} V_e \equiv V_e/\zeta_C \equiv V_e/k_a r_a. \qquad (20)$$

The characteristic *local conversion length (LCL)* $\zeta_C$ is in fact the older of the two, since the parameter "$D$" used in the 1929 Stern, Gossling and Fowler paper [17] is a form of conversion length. This form is used, for example, in Gomer's well-known formula ([18], p.32), where $\zeta_C$ is written as $k_a r_a$, where $r_a$ is the emitter apex radius and $k_a$ is a *shape factor* (also called a "field factor"). However, the characteristic *local voltage-to-barrier-field conversion factor (VCF)* $\beta_{V,C}$ [$=1/\zeta_C$] (as used, for example, in the 1953 Dyke et al. paper [19]) is probably now the more commonly used. Unfortunately, modern LAFE literature tends to use the symbol $\beta$ to denote a real or apparent macroscopic field enhancement factor. To avoid confusion between the various uses of the symbol $\beta$, it is recommended that, in future work, the form involving the LCL $\zeta_C$ should be used to relate $F_C$ to $V_e$.

Note that a local conversion length is not a physical distance (except in very special cases), but is a parameter that reflects both the sharpness of a single-tip emitter and the overall system geometry (sharp emitters that "turn on" at low applied voltages have relatively small conversion lengths).

With LAFEs, the presence of an emitting nanoprotrusion enhances the field at its emitting apex. If a *macroscopic field* $F_M$ is defined in terms of the emission voltage by

$$F_M \equiv V_e/\zeta_M, \qquad (21)$$

where $\zeta_M$ is the *macroscopic conversion length*, then the characteristic value $\gamma_C$ of the *true (electrostatic) macroscopic field enhancement factor (FEF)* is given by

$$\gamma_C \equiv F_C/F_M = \zeta_M/\zeta_C. \qquad (22)$$

TABLE 2. Independent variables, and related auxiliary parameters and equations

| No. | Independent variable | | | Auxiliary parameter | | |
|---|---|---|---|---|---|---|
| | name | symbol | link to | name of parameter | symbol | related formulae |
| | *Theoretical variables* | | | | | |
| T1 | Characteristic local barrier field | $F_C$ | - | - | - | - |
| T2 | Scaled barrier field | $f$ | $F_C$ | Reference field | $F_R$ | $F_C = f F_R$ |
| | *Emission variables* | | | | | |
| T3 | Emission voltage | $V_e$ | $F_C$ | (True) local voltage-to-barrier-field conversion factor (VCF)[a] | $\beta_{V,C}$ | $F_C = \beta_{V,C} V_e$ |
| T4 | Emission voltage | $V_e$ | $F_C$ | (True) local conversion length (LCL) | $\zeta_C$ | $F_C = V_e / \zeta_C$ |
| T5 | True macroscopic field | $F_M$ | $V_e$ | (True) macroscopic conversion length[b] | $\zeta_M$ | $F_M = V_e / \zeta_M$ |
| T6 | True macroscopic field | $F_M$ | $F_C$ | (True) (electrostatic) macroscopic field enhancement factor (FEF) | $\gamma_C$ | $F_C = \gamma_C F_M$ $\gamma_C = \zeta_M / \zeta_C$ |
| | *Measured variables* | | | | | |
| T7 | Measured voltage | $V_m$ | $V_e$ | Voltage ratio | $\Theta$ | $V_e = \Theta V_m$ |
| T8 | Measured voltage | $V_m$ | $F_C$ | Measured-voltage-defined LCL[c] | $\zeta_C^{mvd}$ $= \zeta_C / \Theta$ | $F_C = V_m / \zeta_C^{mvd}$ $F_C = \zeta_C^{-1} \Theta V_m$ |
| T9 | Apparent macroscopic field | $F_A$ | $V_m$ | Macroscopic conversion length | $\zeta_M$ | $F_A = V_m / \zeta_M$ |
| T10 | Apparent macroscopic field | $F_A$ | $V_e$ | No name – not found useful | $\zeta_M \Theta$ | $F_A = V_e / \Theta \zeta_M$ |
| T11 | Apparent macroscopic field | $F_A$ | $F_M$ | Voltage ratio | $\Theta$ | $F_M = \Theta F_A$ |
| T12 | Apparent macroscopic field | $F_A$ | $F_C$ | Apparent-field-defined FEF[d] | $\gamma_C^{afd}$ $= \gamma_C \Theta$ | $F_C = \gamma_C^{afd} F_A$ $F_C = \gamma_C \Theta F_A$ $\gamma_C^{afd} = \zeta_M / \zeta_C^{mvd}$ |

[a] Future use of the parameter $\beta_{V,C}$ is discouraged: use $\zeta_C$ and related formulae instead.
[b] In planar-parallel-plate geometry, $\zeta_M$ is normally taken as equal to the plate separation $d_{sep}$.
[c] Use of the parameter $\zeta_C^{mvd}$ is discouraged: use the combination ($\zeta_C / \Theta$) instead.
[d] Use of the parameter $\gamma_C^{afd}$ is discouraged: use the combination $\gamma_C \Theta$ instead.

When measurements take place in planar-parallel-plate geometry, $F_M$ is often taken as the mean field between the plates and $\zeta_M$ as the separation $d_{sep}$ of the parallel plates, but formulae (21) and (22) are in fact more general than this. The characteristic FEF $\gamma_C$ is a useful LAFE characterization parameter, and in an orthodox emission situation (see below), a $\gamma_C$-value can be extracted from an $i_m(V_m)$-form FN plot, by applying Eq. (22) to the extracted $\zeta_C$-value.

A well known model case is an isolated nanoprotrusion standing on one of a pair of well separated parallel plates. When the nanoprotrusion takes the form of a hemisphere of radius $r_a$ on a cyclindrical post of total height $h$ (including the hemisphere), $\gamma_C$ is given approximately by $0.7h/r$ (e.g., [20]), and the related LCL value is given approximately by

$$\zeta_C \approx (1.4 d_{sep}/h) \cdot r_a. \quad (23)$$

This shows that, in the LAFE case, the shape factor $k_a$ in Gomer's formula $F_a = V/k_a r_a$ is given approximately by $(1.4 d_{\text{sep}}/h)$. The approximation $k_a \sim 5$, as given for STFE geometry by Gomer ([18] p. 32, & [21]) is equivalent to $(d_{\text{sep}}/h) \sim 3.6$. Values of $(d_{\text{sep}}/h)$ used in practical LAFE geometries are usually much larger; this confirms that the approximation $k_a \sim 5$ is not appropriate for LAFEs, as previously pointed out by Edgcombe and Valdrè [22].

### 2.9 Auxiliary parameters – dependent theoretical and emission variables

The dependent theoretical and emission variables normally of interest are shown in Table 3. On letting "L" be "C", $J_C$ and $J_{kC}$ are defined via Eq. (11). The emission current $i_e$ and related parameters are obtained by integrating $J_L$ over the whole surface of the emitter and writing the result in the alternative forms

$$i_e = \int J_L dA \equiv A_n J_C = A_n \lambda_C J_{kC} \equiv A_f J_{kC}, \quad (24)$$

where the *notional emission area* $A_n$ and the *formal emission area* $A_f$ [$\equiv \lambda_C A_n$] are defined via Eq. (24). The reason for introducing both $A_n$ and $A_f$ is the uncertainty in the value of $\lambda_C$. $A_f$ is the area-like parameter that would initially be extracted from a FN plot involving the emission variables, but $A_n$ is the parameter in some existing theory (e.g., [23]) and might correspond more closely to the area seen in a field electron microscope image.

For LAFEs, the *macroscopic current density* $J_M$ is the average ECD taken over the whole *macroscopic area* (or "footprint") $A_M$ of the LAFE, and can be written in the various forms

$$\left.\begin{array}{l} J_M \equiv i_e/A_M = (A_n/A_M)J_C \equiv \alpha_n J_C \\ = \alpha_n \lambda_C J_{kC} \equiv \alpha_f J_{kC} = (A_f/A_M)J_{kC} \end{array}\right\}, \quad (25)$$

where the *notional area efficiency* $\alpha_n$ [$\equiv A_n/A_M$] and the *formal area efficiency* $\alpha_f$ [$\equiv \lambda_C \alpha_n = A_f/A_M$] are defined via Eq. (25). (In earlier work, $\alpha_n$ was denoted by $\alpha_M$, and $\alpha_f$ by $\lambda_M$, and different names were used.)

For the dependent variables, Table 3 includes all the auxiliary parameters and equations currently thought relevant. An important subset consists of those parameters and equations that relate the chosen dependent variable $Y$ to the

TABLE 3. Dependent variables and related auxiliary parameters

| *Dependent variable* | | | *Auxiliary parameter* | | |
|---|---|---|---|---|---|
| name | symbol | link to | name | symbol | definition |
| *Theoretical variables* | | | | | |
| Characteristic kernel current density | $J_{kC}$ | - | - | - | - |
| Characteristic local emission current density (ECD) | $J_C$ | $J_{kC}$ | Characteristic local pre-exponential correction factor | $\lambda_C$ | $= J_C/J_{kC}$ |
| *Emission variables* | | | | | |
| Emission current | $i_e$ | $J_{kC}$ | Formal emission area | $A_f$ | $= i_e/J_{kC}$ |
| Emission current | $i_e$ | $J_C$ | Notional emission area | $A_n$ | $= i_e/J_C$ |
| Macroscopic current density | $J_M$ | $i_e$ | Macroscopic area | $A_M$ | $= i_e/J_M$ |
| Macroscopic current density | $J_M$ | $J_{kC}$ | Formal area efficiency | $\alpha_f$ | $= J_M/J_{kC}$ $= A_f/A_M$ |
| Macroscopic current density | $J_M$ | $J_C$ | Notional area efficiency | $\alpha_n$ | $= J_M/J_C$ $= A_n/A_M$ |

characteristic kernel current density $J_{kC}$: these have the general form

$$Y = c_Y J_{kC}. \qquad (26)$$

For any pair $\{X,Y\}$ of independent and dependent variables, the following relation holds:

$$C_{YX} = c_Y a \phi^{-1} c_X^2. \qquad (27)$$

Universal theory relating to the *measured* dependent variables, when these are different in value from the emission variables, is trickier than it might seem, and is not yet fully developed.

**2.10 The relationship between measured and emission parameters**

Usually, the initial aim of CFE data analysis is to extract (from the measured CFE current-voltage characteristics) values of the parameters equivalent to $c_X$ and (in some cases, where possible) $C_{YX}$ and/or $c_Y$. This requires appreciation of the role of electrical circuit theory. By far the simplest way of dealing with these issues is to work with currents and voltages. By applications of Tables 2 and 3, it can be shown that the $i_e(V_e)$ form of the general FN-type equation can be written

$$i_e = A_f^{GB} a\phi^{-1} (\zeta_C^{-1} V_e)^2 \exp[-v_F^{GB} b\phi^{3/2} \zeta_C / V_e]. \qquad (28)$$

In circuit theory terms, a cold field electron emitter is an electronic device broadly analogous to a pn-junction diode, and has an effective electrical resistance (the *emission resistance* $R_e$) given by:

$$R_e = V_e/i_e = (\phi \zeta_C^2 / A_f^{GB} a V_e) \cdot \exp[v_F^{GB} b\phi^{3/2} \zeta_C / V_e]. \qquad (29)$$

At low emission voltages $R_e$ is very large, but it becomes much smaller as $V_e$ increases.

Fig. 1 is a schematic circuit diagram for CFE measurements. In principle, there may be resistances both in parallel and in series with the emission resistance. Consequently, in principle, the measured current $i_m$ may not be equal to the emission current $i_e$, and the measured voltage $V_m$ may not be equal to the emission voltage $V_e$.

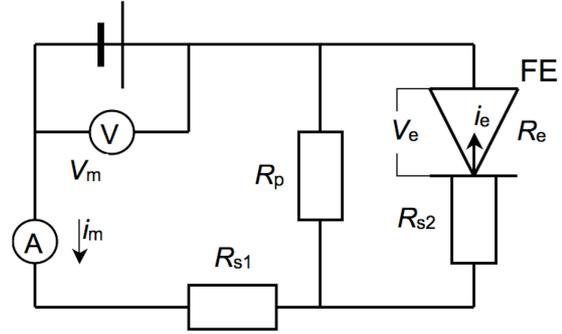

FIG. 1. Schematic electric circuit for measuring the current-voltage characteristics of a field electron emitter.

Resistance in parallel with the emission resistance can usually be made very large by suitable experimental design, hence it is usual and reasonable to assume that $i_m = i_e$. However, a resistance in series with the emission resistance often cannot be eliminated. Let the total series resistance be denoted by $R_s$ [$=R_{s1}+R_{s2}$]. Elementary circuit analysis gives the relationship between $V_e$ and $V_m$ as

$$V_e = V_m - i_m R_s. \qquad (30)$$

Defining a parameter $\Theta$ (the *voltage ratio*) by $V_e = \Theta V_m$ yields

$$\Theta = V_e/V_m = R_e/(R_e + R_s). \qquad (31)$$

Inserting relationship (31) into Eq. (28), and putting $i_m = i_e$, yields the $i_m(V_m)$ form of the general FN-type equation

$$i_m = A_f^{GB} a\phi^{-1} (\zeta_C^{-1} \Theta V_m)^2 \exp[-v_F^{GB} b\phi^{3/2} \zeta_C / \Theta V_m]. \qquad (32)$$

Alternatively, this equation could in principle be written in terms of the *measured-voltage-defined* characteristic local LCL $\zeta_C^{mvd}$, given by

$$\zeta_C^{mvd} \equiv \zeta_C / \Theta. \qquad (33)$$

However, we have concluded that the mathematics is more transparent if the voltage-ratio term $\Theta$ is always shown explicitly in equations, notwithstanding the slightly greater algebraic complexity that this involves.

In modern LAFE literature, it is customary to assume (often incorrectly) that the emission situation is orthodox or elementary, and to state (what the authors probably intend to be) a $J_M(F_M)$-form equation. In very many cases, the published equation is ambiguous and/or incorrect, often in more than one respect. To formulate a correct "macroscopic" equation for non-orthodox situations, it is necessary to define a mathematical *apparent macroscopic field* $F_A$ by

$$F_A \equiv V_m/\zeta_M = V_e/\Theta\zeta_M = F_M/\Theta. \quad (34)$$

Assuming as before that $i_m = i_e$, the related correct general FN-type equation for a non-orthodox situation is the $J_M(F_A)$-form equation

$$J_M = \alpha_f^{GB} a\phi^{-1}(\gamma_C \Theta F_A)^2 \exp[-\nu_F^{GB} b\phi^{3/2}/\gamma_C \Theta F_A]. \quad (35)$$

Note that the formal area efficiency $\alpha_f^{GB}$ and the voltage ratio $\Theta$ appear in this equation.

In our view, the relationship between $F_A$ and the true macroscopic field $F_M$ is intellectually more awkward to deal with than the corresponding relationship between $V_m$ and $V_e$, especially for students. Hence, because theory involving $V_m$ and $V_e$ is more straightforward, our firm view is that FN-plot analysis is best carried out on $i_m(V_m)$-form FN plots based on the raw experimental data (or, failing this, on $J_M(V_m)$-form FN plots). Discussion below deals with $i_m(V_m)$-form FN plots.

## 3. Current-voltage data analysis using Fowler-Nordheim plots

### 3.1 Introduction

Given Eqs. (16) and (18), a quantity $L(X^{-1})$ can be defined by

$$\begin{aligned} L(X^{-1}) &\equiv \ln\{Y/X^2\} \\ &= \ln\{C_{YX}\} - B_X/X \\ &= \ln\{C_{YX}\} - \nu_F b\phi^{3/2}/(c_X X) \end{aligned} \quad (36)$$

A FN-type equation written in this way is said to be *"written in FN coordinates"*.

A *Fowler-Nordheim plot (FN plot)* is a plot of $L(X^{-1})$ versus $X^{-1}$. General features of FN-plot theory are common to all *forms* of FN plot (i.e., such general features do not depend on the particular independent and dependent variables used), and are best discussed using the universal FN-type equation above. This has been done elsewhere [14]. A summary of the most relevant part of this treatment is presented here. For notational simplicity, the superscript "GB" is now dropped, but relevant quantities still apply to a general barrier unless indicated otherwise.

Except in the case of the ET barrier, $\nu_F$ (and hence $B_X$) are functions of $X^{-1}$. The parameters $c_X$, $c_Y$ and $C_{YX}$, too, will often be functions of $X^{-1}$. Hence, theoretical FN plots are expected to be curved (though for metal emitters this curvature is so slight as to be barely noticeable). At any given value of the horizontal-axis value $X^{-1}$, the slope $S_{YX}(X^{-1})$ of a theoretical FN plot is given by

$$\begin{aligned} S_{YX}(X^{-1}) &= dL/d(X^{-1}) \\ &= [d\ln C_{YX}/d(X^{-1}) - X^{-1}\nu_F d(b\phi^{3/2}c_X^{-1})/d(X^{-1})] \\ &\quad - [(b\phi^{3/2}c_X^{-1})\{\nu_F + X^{-1}d\nu_F/d(X^{-1})\}] \end{aligned}$$
(37)

By introducing the *slope correction function* $\sigma_{YX}(X^{-1})$ defined by

$$\sigma_{YX}(X^{-1}) \equiv [dL/d(X^{-1})]/(b\phi^{3/2}c_X^{-1}), \quad (38)$$

the slope $S_{YX}(X^{-1})$ can be written in the much simpler form

$$S^{tan} = S_{YX} = -\sigma_{YX} \cdot (b\phi^{3/2}c_X^{-1}) = \sigma_{YX} \cdot S_{YX}^{el}, \quad (39)$$

where $S_{YX}^{el}$ [$= -b\phi^{3/2}c_X^{-1}$] is the slope of the plot for the corresponding elementary FN-type equation. For notational simplicity, we no longer explicitly show the dependence on $X^{-1}$. Obviously, a tangent to this theoretical FN plot, taken at abscissa-value $X^{-1}$, also has slope $S^{tan}$ as given by Eq. (39).

It can be shown [14] that this tangent intersects the $L$-axis at the value $\ln\{I^{tan}(X^{-1})\}$ given via

$$I^{\tan}(X^{-1}) = \rho_{YX}(X^{-1}) \cdot C_{YX}(X^{-1}), \tag{40}$$

where the *intercept correction function* $\rho_{YX}(X^{-1})$ is given by

$$\left.\begin{aligned}\ln\{\rho_{YX}\} &= [\sigma_{YX} - \nu_F] \cdot [b\phi^{3/2}/c_X X] \\ &= [\sigma_{YX} - \nu_F] \cdot G_F^{ET}\end{aligned}\right\}, \tag{41}$$

where $G_F^{ET} \equiv b\phi^{3/2}/F_C$. Apart from the universal constant $b$, all parameters in this formula are or can be functions of $X^{-1}$, but this is not shown explicitly.

Note that, in this paper, all correction functions $\rho$ are the new type of intercept correction function introduced in Ref. [14], rather than the older type of intercept correction function used in pre-2012 papers. The subscripts $YX$ are included as a reminder that, in principle, the forms of the correction functions may depend on which specific variables are represented by $X$ and $Y$.

Obviously, experimental data can also be plotted on a FN plot. Often, but not always, an experimental FN plot is either a nearly straight line, or basically breaks into two nearly straight segments. A nearly straight FN plot or segment is usually analysed by fitting to it a straight line with (usually negative) slope $S_X^{fit}$, intercept $I_{YX}^{fit}$, and equation

$$Y = \ln\{I_{YX}^{fit}\} + S_X^{fit}/X. \tag{42}$$

Normally, the task is to extract estimates of $c_X$ (and sometimes $C_{YX}$ and/or $c_Y$) from the measured values $S_X^{fit}$ and $I_{YX}^{fit}$. For non-metals this can be far more complicated than has been generally realized. The abstract principles involved (set out below) are becoming clear, but our detailed understanding of how to do this reliably in real emission situations is still very much under development.

### 3.2 The tangent method

Although other methods of FN plot analysis exist, the most flexible method is the *tangent method*. In this method it is assumed that the straight line fitted to the experimental data can be modelled by a tangent to the theoretical plot, taken at some specific value $X_t^{-1}$ called the *fitting value*. Functions evaluated at the fitting value are subscripted "t", and the name "factor" (rather than "function") is used to indicate the value thus obtained. Thus, the tangent to the theoretical plot, taken at the abscissa value $X_t^{-1}$, can be written as the straight line

$$\left.\begin{aligned}Y &= \ln\{I^{\tan}\} + S^{\tan}/X \\ &= \ln\{\rho_{YX,t}C_{YX,t}\} - \sigma_{YX,t}b\phi^{3/2}/c_{X,t}X\end{aligned}\right\}, \tag{43}$$

where $\sigma_{YX,t}$, $\rho_{YX,t}$, $C_{YX,t}$ are $c_{X,t}$ are values taken at the fitting value $X_t$.

This assumption that the fitted line can be modelled by a tangent is not exactly true, because the fitted line is in principle a chord to the theoretical FN plot. In principle, a *chord correction* [24] could be made, but this is difficult to do exactly, and there is no evidence that making a chord correction significantly affects final extracted values. (Other uncertainties are nearly always much greater than the uncertainty associated with neglecting the chord correction.)

In principle, if no chord correction is made, the data-analysis procedure is then to identify Eq. (42) with Eq. (43) and extract values of $c_{X,t}$, $C_{YX,t}$ and $c_{Y,t}$ using the formulae

$$c_{X,t}^{extr} = -\sigma_{YX,t}b\phi^{3/2}/S_X^{fit}, \tag{44}$$

$$C_{YX,t}^{extr} = I_{YX}^{fit}/\rho_{YX,t}, \tag{45}$$

$$\left.\begin{aligned}c_{Y,t}^{extr} &= C_{YX,t}^{extr}/\{a\phi^{-1}(c_{X,t}^{extr})^2\} \\ &= \{I_{YX}^{fit}(S_X^{fit})^2\}/[(ab^2\phi^2)(\rho_{YX,t}\sigma_{YX,t}^2)]\end{aligned}\right\}. \tag{46}$$

If a chord correction is made, then $I_{YX}^{fit}$ in formulae (45) and (46) is replaced [24] by $I^{corr}$, where

$$I^{corr} = \rho^{chord}I_{YX}^{fit}, \tag{47}$$

where $\rho^{chord}$ is a *chord correction factor*. As indicated above, expected values for $\rho^{chord}$ are not reliably known; however, an approximate estimate can be obtained from the work of Spindt et al. [25], who fitted a chord to a plot of $\nu(y)$ vs $y$ (see their Fig. 5). It has been shown [24] that their result is equivalent to taking $\rho^{chord} \approx 1.2$.

The correction factors $\sigma_{YX,t}$ and $\rho_{YX,t}$ cannot be measured, but have to be estimated theoretically, using specific (physically plausible) mathematical assumptions about the forms of $v_F$ and $C_{YX}$, and an estimate of the value of $X_t^{-1}$. Herein lie two of the main difficulties of FN plot analysis.

In the universal FN-type Eq. (16), all of $v_F^{GB}$, $\phi$, $c_X$, $c_Y$ and $C_{YX}$ can in principle be functions of $X^{-1}$. Hence, the detailed forms of Eq. (37), and hence Eqs. (38) and (41) can in principle be very complicated, involving many individual terms, many of which have never been carefully investigated. No general approach seems practicable. These formulae can be applied successfully in the so-called orthodox emission situation (see below), where $\phi$, $c_X$, $c_Y$ and $C_{YX}$ are assumed constant and the mathematical forms of $\sigma_{YX}$ and $\rho_{YX}$ are well known, but other situations are problematical. To progress scientifically, it looks necessary to proceed in a series of focused investigations, each of which involves specific mathematical approximations that allow some specific physical effect or effects to be explored. One initial line of investigation has been into the effects of barrier form, as discussed below.

One also has to determine the fitting value, $X_t^{-1}$. The first estimate is always the mid-point of the range of $X^{-1}$-values covered by the experimental data being analyzed. However, when the curvature of a FN plot is non-uniform (which is usually the case), then the mid-range value is probably not the best choice, and the error in the extracted result is in principle slightly increased (especially for $A_f$ and $\alpha_f$). For example, in the orthodox emission situation, a theoretical plot based on a Schottky-Nordheim barrier has greater curvature on the left-hand side (low $X^{-1}$-value side), and the best choice of $X_t^{-1}$ is somewhat to the left of the mid-range value, as demonstrated in Ref. [26]. The issue of the best choices of $X_t^{-1}$ for FN plots related to non-orthodox situations has not yet been systematically investigated.

In cases where the correction functions are slowly varying functions of $X^{-1}$, the exact choice of fitting value is not important. Thus, in orthodox and "nearly orthodox" emission situations, the slope correction factor can usually be adequately approximated as $s_t \equiv s(f_t) \approx 0.95$, where $s$ is the well-known slope correction function for the SN barrier (e.g., [11, 16]). However, in cases where FN plots are obviously curved, and hence $\sigma_{YX}$ must be varying relatively rapidly, the choice of $X_t^{-1}$ is expected to be important. More research is needed on this issue.

### 3.3 Analysis of $i_m(V_m)$-form FN plots

As indicated above, a FN plot of the form $[\ln\{i_m/V_m^2\}$ vs $1/V_m\}$ is called here an $i_m(V_m)$-*form* FN plot. For convenience, the subscript "m" or "mm" (rather than the variables themselves) will be used to label parameters belonging to an $i_m(V_m)$-form FN plot. For such a plot, $c_X$ becomes $c_m = \Theta \zeta_C^{-1}$; thus, from Eq. (44), an extracted value $\zeta_C^{extr}$ of the true local conversion length $\zeta_C$ can in principle be derived from the FN-plot slope using

$$\left.\begin{array}{rl}(\zeta_C^{extr})^{-1} &= -(\sigma_{mm,t}/\Theta_t) \cdot (b\phi^{3/2})/S_{mm}^{fit} \\ &\equiv -\sigma_{SR} b\phi^{3/2}/S_{mm}^{fit}\end{array}\right\}, \quad (48)$$

where $\Theta_t$ is the voltage ratio at the $V_m$-value where the tangent is taken, and $\sigma_{SR}$ is the *effective slope correction factor for the series resistance (SR) situation.* This parameter $\sigma_{SR}$ is defined by Eq. (48) and given in terms of $\sigma_{mm,t}$ by

$$\sigma_{SR} = \sigma_{mm,t}/\Theta_t. \quad (49)$$

For analogy with what is done in some existing literature, we also introduce an $i_m(V_m)$-*form slope characterisation parameter* $\zeta_C^{app}$ (or "apparent LCL") defined via

$$(\zeta_C^{app})^{-1} \equiv -b\phi^{3/2}/S_{mm}^{fit}. \quad (50)$$

The correct extracted value $\zeta_C^{extr}$ of the true LCL $\zeta_C$ is related to this via

$$(\zeta_C^{\text{extr}})^{-1} = \sigma_{\text{SR}} \cdot (\zeta_C^{\text{app}})^{-1}, \tag{51}$$

and, using Eq. (22), the corresponding extracted value $\gamma_C^{\text{extr}}$ of the true FEF $\gamma_C$ is related to $\zeta_C^{\text{app}}$ via

$$\gamma_C^{\text{extr}} = \sigma_{\text{SR}} \cdot [\zeta_M / \zeta_C^{\text{app}}]. \tag{52}$$

The reciprocals of $\zeta_C$ and $\zeta_C^{\text{app}}$ are used above, rather than the parameters themselves, in order to make the relevant formulae look similar to FEF-related formulae existing in the literature and discussed below.

A common (but unfortunate) literature approach is to pre-convert the experimental data, in effect by using the formulae $F_A = V_m/\zeta_M$, $J_M = i_m/A_M$, and make a $J_M(F_A)$-form FN plot. Let this have fitted slope $S_{\text{MA}}^{\text{fit}}$. In the elementary data-analysis approach nearly always used, a *slope characterization parameter* (or *apparent FEF*, or *pseudo-FEF*), denoted here by $\beta^{\text{app}}$, is then derived from the formula

$$\beta^{\text{app}} = -b\phi^{3/2}/S_{\text{MA}}^{\text{fit}}. \tag{53}$$

Since $S_{\text{MA}}^{\text{fit}}/S_{\text{mm}}^{\text{fit}} = \zeta_M^{-1}$, it follows from Eqs. (50) and (53) that

$$(\zeta_C^{\text{app}})^{-1} = \beta^{\text{app}}/\zeta_M, \tag{54}$$

and hence, from Eq. (52), that the correct extracted value $\gamma_C^{\text{extr}}$ of the true electrostatic FEF is obtained from

$$\gamma_C^{\text{extr}} = \sigma_{\text{SR}} \cdot \beta^{\text{app}}. \tag{55}$$

A serious weakness of much modern LAFE literature is that it uses the same symbol "$\beta$" for both of the quantities denoted here by $\gamma_C$ and $\beta^{\text{app}}$. This hides the existence of $\sigma_{\text{SR}}$, and is equivalent to taking $\sigma_{\text{SR}}=1$. This can be a very poor approximation when series resistance or "saturation" is present, as it is known [13] that in these circumstances $\sigma_{\text{SR}}$ can be significantly less than unity. Consequently, many FEF-values reported in the literature are in fact spuriously large [13].

We make the trivial point that, although formula (55) looks marginally simpler than formula (52), this is deceptive. In order to construct a $J_M(F_A)$-form FN plot, it is necessary to divide every measured voltage ($V_m$) value by $\zeta_M$ (in practice usually by $d_{\text{sep}}$). It is less work to directly use the $V_m$-values in the FN plot, and multiply the fitted slope value, or the corresponding slope characterization parameter value, by $\zeta_M$ (in practice usually by $d_{\text{sep}}$).

Physically, $\zeta_C$ should often be a relatively well-defined quantity, with a relatively well-defined value. Extraction of a reliable value is straightforward in "orthodox" emission situations, as defined below, in which (by definition) no series resistance or other complications are present (so $\Theta=1$), and $\sigma_{\text{mm,t}}$ and $\sigma_{\text{SR}}$ become given by $s_t \approx 0.95$.

However, the situation is different when significant series resistance is present. Extraction of a reliable value for $\zeta_C$, using Eq. (47) or Eq. (50), then needs good estimates of the values of both the voltage ratio $\Theta_t$ and the slope correction factor $\sigma_{\text{mm,t}}$ (which in this case may include terms derived by differentiation with respect to $\Theta$). A further difficulty is that the series resistance $R_s$ may be current-dependent, with the nature of this dependence depending on the detailed physical nature of the conducting path between the voltage generator and the tip emitting region. At present, there is very little systematic practical knowledge about the likely behaviour and value of $R_s$ for conducting paths that are not exclusively metallic; consequently, very little empirical knowledge exists about likely values of $\Theta_t$ or $\sigma_{\text{mm,t}}$. These difficulties mean that, when significant series-resistance effects occur, it is currently impossible to extract reliable $\zeta_C$-values or reliable $\gamma_C$-values from an $i_m(V_m)$-form plot (or reliable $\gamma_C$ values directly from an $J_M(F_A)$-form plot). Discussion of this is continued in Section 4.

### 3.4 The barrier-effects-only approximations

Even if one makes the assumption that no series resistance is present (hence $\Theta = 1$, and

d$\Theta$/d$V_m$=0), there remain several effects (for example, field dependent changes in emission-system geometry, when a carbon nanotube is pulled upwards by Maxwell stress) that can in principle create $V_e$-dependence in $B_X$ and/or $C_{YX}$, and some of these have never been investigated in detail. What is now needed is systematic investigation of the various possibilities, where this is practicable.

With Eqs. (38) and (41), the *barrier-effects-only approximation* (previously called the "basic approximation" [14]) is to take into account only those terms that relate to the direct dependence on $X^{-1}$ and the dependence of $\nu_F$ on $X^{-1}$. This approximation disregards all terms in the first square bracket in Eq. (37). In the barrier-effects-only approximation, the various independent theoretical and emission variables are linearly related, and the general correction functions $\sigma_{YX}$ and $\rho_{YX}$ become given by *slope and intercept correction functions* $\sigma^B$ and $\rho^B$ defined by the formulae [14]

$$\left.\begin{array}{rl}\sigma^B & \equiv \nu_F - X_C d\nu_F/dX_C \\ & = \nu_F - F_C d\nu_F/dF_C\end{array}\right\}, \quad (56)$$

$$\left.\begin{array}{rl}\ln\{\rho^B\} & \equiv [\sigma^B - \nu_F]\cdot(b\phi^{3/2}/F_C) \\ & = [\sigma^B - \nu_F]\cdot G_F^{ET}\end{array}\right\}, \quad (57)$$

where $X$ here is one of the theoretical or emission variables.

Eqs. (56) and (57) apply to a barrier of any mathematical form. As indicated above, the so-called *orthodox approximation* involves, *in addition*, the assumption that the barrier is a Schottky-Nordheim barrier. In this case, $\nu_F$ becomes $\nu_F^{SN}$ and is given by the relevant particular value $v(f)$ of the principal SN-barrier function $v(l')$ [11, 15], $\sigma^B$ becomes given by the SN-barrier function $s$ [11], $\rho^B$ becomes given by the SN-barrier function $r_{2012}$ discussed in Ref. [14], and the defining equations reduce to [11]:

$$s = v - f dv/df, \quad (58)$$

$$\left.\begin{array}{rl}\ln\{r_{2012}\} & = [s-v]\cdot G_F^{ET} \\ & = -(fdv/df)G_F^{ET} = \eta u\end{array}\right\}. \quad (59)$$

where all relevant parameters denote characteristic values, but this is not shown explicitly. Here, $u$ is the SN-barrier function defined by $u = -dv/df$ [11] (see Appendix C).

The so-called *elementary approximation* (much used in modern LAFE literature) involves, *in addition to* the assumptions made at the start of this Section, the assumption that the barrier is exactly triangular. In this case $\sigma^B$=1 and $\rho^B$=1.

### 3.5 The orthodox data-analysis approach

The *orthodox approach* to FN-plot analysis is based on making a set of physical and mathematical assumptions, about the physical measurement situation and about the theory of emission. This *orthodox emission situation* is defined formally in Appendix B. In such a situation, FN-plot analysis is straightforward. No real emission situation is "exactly orthodox", but many real situations are "very nearly orthodox", and the various assumptions made are adequately valid.

This "orthodox approach" is a development of earlier methods, in particular those used by Charbonnier and Martin [27] and Spindt et al. [25]; these were based on the Murphy-Good FN-type equation [12], which assumes emission through a SN barrier, but approximates $\lambda_L^{SN}$ as equal to a mathematical pre-factor sometimes denoted by $t_F^{-2}$ (see [11]). Exact analytical forms and numerical values have long been known for the SN-barrier functions (see [11]), but the existence of simple approximations for $v$ and $s$ has been key to their application in CFE data analysis. Many such approximations have been proposed; some of these are listed in [24]. In recent years, significant steps forwards have been: (a) the realization that the natural physical variable for the argument of $v$ is the scaled barrier field $f$, rather than the Nordheim parameter $y$ [=$f^{1/2}$] previously used; and (b) the discovery of a good, simple, accurate algebraic approximation for *v*(*f*), namely Eq. (12) above. It has been shown [24] that, on average over the range $0 \leq f \leq 1$, Eq. (12) is more accurate than any other approximation of equivalent complexity. Related approximate expressions can be given [11] for relevant SN-barrier functions, including *u*(*f*), *s*(*f*) and $r_{2012}(f)$, and also have good accuracy. These expressions are given in Appendix C.

By using simulated input data for electron escape through an SN barrier, and these simple expressions for *s*(*f*) and $r_{2012}(f)$, it has been demonstrated that (in an orthodox emission situation) application of the tangent method leads to accurate extraction of emission characterization parameters, in particular the field enhancement factor and formal emission area [14].

## 3.6 Correction factors for other barrier shapes

Many earlier calculations of local emission current density have included calculations of the barrier form correction factor $v_F$, for a variety of emitter shapes and related barrier forms. However, there have been few calculations of slope and intercept correction factors. Recent explorations [28, 29] have generated the following conclusions.

(1) For planar emitters, the precise form of the model used for the exchange-and-correlation (XC) contribution has relatively little effect on predicted values of $\sigma^B$ and $\ln\rho^B$ [28].

(2) For a spherical emitter, the barrier (if XC effects are disregarded) is the so-called *Coulomb barrier*, well known in nuclear physics. This barrier has an analytical solution for $v_F$ from which estimates can be derived for $\sigma^B$ and $\ln\rho^B$. Theoretically predicted FN plots for the Coulomb barrier can be significantly curved, particularly for low-radius emitters [29].

(3) For a spherical emitter, as with a planar emitter, the inclusion of an XC term has a significant effect of the values of $v_F$, $\sigma^B$ and $\ln\{\rho^B\}$. However, provided the sphere radius is "not too small", there is relatively little difference between the results of using Schottky's planar image-PE approximation to model XC effects and the results of using the spherical image-PE approximation [29].

(4) For a spherical emitter of small radius, the electrostatic term ceases to be a valid approximation for electrostatic effects associated with a real single-tip-geometry emitter, because the influence of the emitter shank increases as the tip radius decreases. For a real small-apex radius emitter, the sphere-on-orthogonal cone (SOC) model [19] represents the electrostatics better. Preliminary investigations [29], illustrated in Fig. 2, show that, as expected, the results for the two models diverge as apex radius decreases.

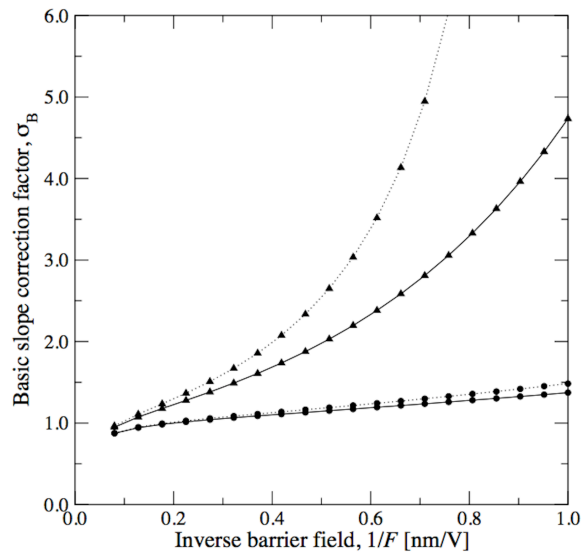

FIG. 2. Plots illustrating how the slope correction function $\sigma^B$ varies as a function of inverse barrier field $1/F$. Dotted lines are for a spherical emitter; continuous lines are for an emitter modelled using the sphere-on-orthogonal-cone (SOC) model [19]. Circles are for apex radius $r_a = 20$ nm; triangles are for $r_a = 5$ nm. SOC-model parameters used were: $\phi = 4.50$ eV, $n=0.1$, $r_a$ as indicated, $a = r_a/4$.

Unfortunately, evidence has recently come to light [30] that the quasi-classical quantum-mechanics used in existing analyses of CFE from non-planar emitters is likely to be valid only when the Schrödinger equation separates in Cartesian coordinates, which, obviously, is not the case for non-planar emitters. Hence, most or all existing treatments of field electron emission from non-planar emitters may need adjustment. This is an active topic of research, but it is not yet clear what detailed form the theoretical adjustment should take. Consequently, we do not give further examples of previous (unadjusted) results here.

This leaves us with the situation that (with the exception of a few special cases) the only emitters where we can be sure that the results of FN plot analysis are strictly valid are those that comply adequately with the conditions for orthodox emission. A test for identifying emitters that do *not* comply now exists [13], and is described next.

### 3.7. A test for lack of orthodoxy

For most of the last 50 years, FN-plot analysis has used either the orthodox approach (using either the tangent method or a simplified version of it), or an *elementary approach* based on the elementary approximation defined above.

However, the assumptions of the orthodox emission situation exclude many complications that can occur in real situations. The excluded complications include significant effects resulting from: series resistance; leakage currents; patch fields; field emitted vacuum space-charge; current-induced changes in emitter temperature; field penetration and band-bending; quantum confinement; and field-related changes in emitter geometry, emission area and/or local work-function. These complications can affect measured current-voltage characteristics.

The test for "lack of orthodoxy" (i.e., whether the measured characteristics are incompatible with the orthodox emission hypothesis) involves extracting $f$-values from an experimental FN plot, using the formula

$$f^{\text{extr}} = -s_t \eta / [S^{\text{fit}} \cdot (X^{-1})^{\text{expt}}], \qquad (60)$$

with $s_t \approx 0.95$. For simplicity, we no longer show either universal variable $X$, $Y$ as a subscript. The test, however, applies to any form of FN plot. By using Eq. (60), $f$-values can be extracted that correspond to the range of $X$ values measured.

For any given $\phi$-value, a set of four indicative boundary $f$-values $\{f_{\text{lb}}, f_{\text{low}}, f_{\text{up}}, f_{\text{ub}}\}$ can be defined as discussed in Ref. [13]. For $\phi$=4.50 eV, these values are {0.10, 0.15, 0.45, 0.75}; for other $\phi$-values, boundary $f$-values can be derived from Table 2 in Ref. [13]. The range of extracted $f$-values is then compared with these boundary values. One of the following three situations then applies. (1) If the extracted range is totally inside the range $\{f_{\text{low}} < f^{\text{extr}} < f_{\text{up}}\}$ of "apparently reasonable" $f$-values, then the orthodoxy test is passed. (2) If any part of the extracted range is "clearly unreasonable" because $f^{\text{extr}} < f_{\text{lb}}$ or $f^{\text{extr}} > f_{\text{ub}}$, then the orthodoxy test is "clearly failed". (3) If some part of the extracted range is outside the "clearly reasonable" range, but not in either of the "clearly unreasonable" ranges, then "further investigation is needed".

If case (1) applies, then characterization data extracted from the FN plot (in particular, the apparent FEF-value) can be treated as reliable. If case (2) applies, then extracted characterization data are almost certainly spurious. If case (3) applies, then extracted characterization data are unreliable, and are more likely than not to be spurious.

### 4. Analysis of measured current-voltage data when significant series resistance exists

Although series resistance in the conducting path from the high-voltage generator to the emitting region at the tip apex is not the only possible cause of orthodoxy-test failure, it is currently thought to be the commonest cause. As just indicated, when the orthodoxy test is failed, then the usual "orthodox" and "elementary" methods of extracting characterization parameters from FN plots will generate spurious results. In Ref. [13], the orthodoxy test described above was applied to a small sample of 19 published FN plots, taken from emitters fabricated from various non-metals. Approximately 40% of these failed the test, indicating that the associated published FEF-values are spuriously large. If this sample is representative of the literature as a whole (which may or may not be the case), then one might expect that there are many hundreds of published field emission papers that report FEF-values that are in fact spuriously large.

Consequently, it was argued in Ref. [13] that (certainly until such time as we better understand any systematic trends involved) the orthodoxy test should always be applied to FN plots taken from non-metallic emitters, and the results of the test should be published alongside any published FN plot.

Obviously, better methods of data analysis for non-orthodox emission situations are also needed. This Section describes progress with explorations into several methods of extracting more reliable characterization data when the presence of series resistance is thought (or hypothesised) to be the only significant problem.

### 4.1 Analysis via a slope correction factor, for the Schottky-Nordheim barrier

We first report an investigation [31] that originally aimed to estimate the effective slope correction factor $\sigma_{SR}$ [$=\sigma_{mm,t}/\Theta$], assuming constant series resistance and emission through a SN barrier. This scenario has analytical solutions, since all relevant quantities can be expressed as functions of chosen constant-values and a single variable, namely the (true) scaled barrier field $f$. This makes the scenario particularly suitable for analysis by computer algebra packages, in particular the MAPLE™ package used by one of us [JHBD].

The emission voltage $V_e$, emission current $i_e$ and emission resistance $R_e$ are given by

$$V_e = \zeta_C F_R f, \tag{61}$$

$$i_e = A_f \theta f^2 \exp[-v(f) \cdot \eta/f], \tag{62}$$

$$R_e = V_e/i_e. \tag{63}$$

The measured current $i_m$ can be taken as equal to $i_e$, and the measured voltage $V_m$ is given by

$$V_m = V_e + i_e R_s. \tag{64}$$

Expressions for $dV_m$ in terms of $df$, and for $di_m$ in terms of $df$ can be obtained via Eqs. (64) and (62), respectively, and the slope $S_{mm}$ of the $i_m(V_m)$-form FN plot can be evaluated from

$$S_{mm} = V_m \left[ 2 - i_m^{-1} V_m (di_m/dV_m) \right]. \tag{65}$$

From this, predicted values of the effective slope correction function in the series-resistance situation ($\sigma_{SR}$) can be obtained from

$$\sigma_{SR} = S_{mm}/(b\phi^{3/2}\zeta_C). \tag{66}$$

Fig. 3a shows the predicted $i_m(V_m)$-form FN plots, for the parameter values $\phi$=4.50 eV (hence $F_R$= 14.06 V/nm) and $\zeta_C$ = 100 nm, and for various values of the product ($A_f R_s$). Fig. 3b shows a plot of $\sigma_{SR}$ vs $1/V_m$ for the value $A_f R_s = 10^{-1}$ Ω m². Note that the predicted FN plots "turn over" at the left-hand side.

These predicted FN plots do not well resemble experimental FN plots found in the literature for materials that are thought to fail the orthodoxy test because of series resistance. Further, the parameter $\sigma_{SR}$ varies significantly with $1/V_m$, which makes choice of a fitting value difficult. At present, the precise reasons for this non-agreement of predicted and measured FN plot shapes are not clear. In many cases the most plausible reason might be current-dependence in $R_s$, but other possibilities need thinking about. These might include field-dependent changes in emitter geometry or some anomalous effect associated with the fact (with LAFEs) the current is drawn from many individual emitting tips, rather than a single tip.

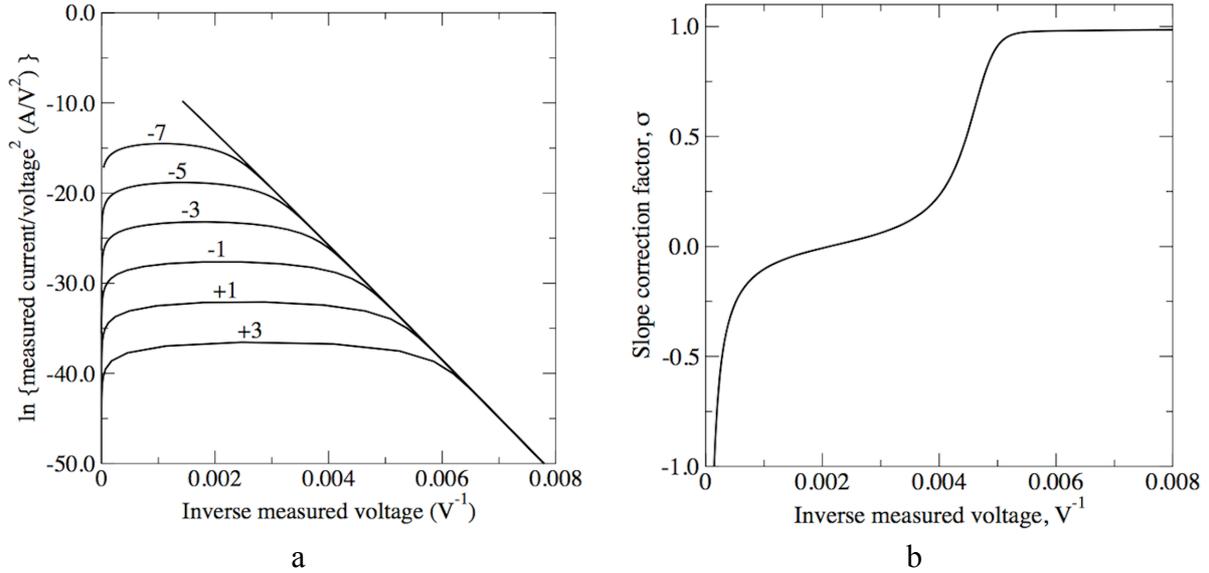

FIG. 3. To illustrate the results of simulating circuit performance when a high resistance $R_s$ is placed in series with a field electron emitter (with formal emission area $A_f$) operating in accordance with the new-standard FN-type equation. (a) $i_m(V_m)$-form FN-plots, for various values of the product $A_f R_s$. A curve marked "$N$" corresponds to an $A_f R_s$-value of $10^N$ Ω m². (b) Plot showing the effective slope correction factor $\sigma_{SR}$ as a function of inverse measured voltage $V_m^{-1}$, for $A_f R_s = 10^{-1}$ Ω m².

Provisional conclusions are that future theoretical research needs to explore the consequences, for predicted FN plots, of current-dependence in the series resistance, but that other methods of dealing with series resistance need further exploration, and could be more effective in the short term.

**4.2 Analysis via simulation of constant series-resistance effects**

The most obvious alternative method involves the assumptions that the series resistance is, in fact, constant, and that an $i_e(V_e)$-form FN plot ought to be a nearly straight line. This method has had longest use in the context of electronic random-access memory (e.g., [32]), but related thinking has also been used in the context of LAFEs (see [33-35]). Some criterion (which may be statistical or empirical) is needed for assessing the straightness of the FN plot. The method then works as follows. A value is assumed for $R_s$. For each of a range of values of $i_m$ ($=i_e$), a value $V_e$ is obtained from Eq. (19), and an $i_e(V_e)$-form FN plot is drawn. This procedure is carried out for a range of values of $R_s$, and the straightest resulting plot is selected as the definitive plot. Emitter characterization parameters are then derived from this "definitive" $i_e(V_e)$-form plot, by a conventional orthodox or elementary method.

This procedure, as applied in the context of electronic random access memory, is well described by Miranda [32]. There is scope for further theoretical investigation of how best to apply it in the context of LAFEs, but it should be remembered that the assumption of constant series resistance may be more plausible for electronic random access memory devices than for LAFEs.

**4.3 Phenomenological adjustment**

A further alternative method is *phenomenological adjustment* of the extracted slope characterization parameter, which we denote here by $c_X^{app}$. In practice, $c_X^{app}$ would normally be one of the parameters $(\zeta_C^{app})^{-1}$ or $\beta^{app}$ discussed earlier.

The method of "phenomenological adjustment" [31] assumes that emission in fact takes place through a SN barrier, and takes place over the same range of $f$-values as does orthodox emission. A value $(1/f^{orth})_{mid}$ is assumed for the typical mid-range value of $1/f$ for practical orthodox field emitters; a value around 4 seems appropriate. This value is then compared with the "experimental" value $(1/f^{\text{extr}})_{\text{mid}}$ extracted, via Eq. (60), for the $X^{-1}$-value that corresponds to the mid-point of the range of experimental $X^{-1}$ values used in the experiments, and an *adjustment factor* $\omega_{\text{adjust}}$ is obtained from

$$\omega_{\text{adjust}} = (1/f^{\text{extr}})_{\text{mid}} / (1/f^{\text{orth}})_{\text{mid}}. \tag{67}$$

Non-orthodox emitters often fail the orthodoxy test because their extracted $f$-values are too high ($1/f$ too low); in this case the value of $\omega_{\text{adjust}}$ will be less than unity.

The method assumes that a phenomenological estimate $c_X^{\text{est}}$ of the true value of the auxiliary parameter $c_X$ can be obtained from the extracted slope characterization parameter $c_X^{\text{app}}$, by using $\omega_{\text{adjust}}$ as a "phenomenological estimate" of the slope correction factor $\sigma_{\text{SR}}$ and hence that

$$c_X^{\text{est}} \sim \omega_{\text{adjust}} \cdot c_X^{\text{app}}. \tag{68}$$

The method will work for FEFs, VCFs and (the reciprocals of) LCLs. In the case of FEFs, Eq. (68) becomes

$$\gamma_C^{\text{est}} \sim \omega_{\text{adjust}} \cdot \beta^{\text{app}}, \tag{69}$$

where $\beta^{\text{app}}$ is the slope characterization parameter (apparent FEF) as defined by Eq. (53). (In the literature, $\beta^{\text{app}}$ is usually denoted simply by $\beta$, and is usually simply called a FEF.) $\gamma_C^{\text{est}}$ is the estimated ("ansatz-adjusted") value of the true characteristic FEF for the emitter concerned, as obtained by phenomenological adjustment.

As a specific (worst-case) example, consider data-entry 20 in Table 4 in Ref. [13]. This entry relates to the high-field part of a FN plot for what is described as a "flexible $SnO_2$ nanoshuttle", and a FEF-value of around 130 000 is reported. The $f$-value range derived as part of the orthodoxy test is 5.6 to 33.2. This corresponds to a mid-$(1/f)$-range value of $(1/f^{\text{extr}})_{\text{mid}} = 0.104$, and (if we take $(1/f^{\text{orth}})_{\text{mid}}$ as 4, as suggested above) to an adjustment-factor $\omega_{\text{adjust}}$ of 0.026. Thus the adjusted FEF-value $\gamma_C^{\text{est}}$ is around 3400.

The estimates derived via the above procedure should not be taken as scientifically valid numerical estimates of $c_X$-values, but they should have sufficient qualitative validity to be useful in technological contexts. In particular, this adjustment has the effect of reducing published FEF-values for materials that fail the orthodoxy test, and this may be of use in literature searches for materials with especially high values of true (electrostatic) macroscopic FEF. However, it is emphasized that this method currently has only the limited scientific basis set out here. A need exists for its degree of validity to be investigated by appropriate simulations, but currently this is difficult, because in many or most cases we do not currently know how to correctly model the effect that is presumed to be responsible for the problem in the first place.

## 5. Discussion

Hopefully, this paper demonstrates that useful progress is being made, both in giving a systematic scientific description of Fowler-Nordheim theory, and in establishing how to correctly analyze FN plots taken from materials that are not good conductors. Particular recent advances, discussed above, have been the incorporation of the voltage ratio $\Theta$ into equations, and the idea of phenomenological adjustment. However, much more remains to be done. The following seem immediately useful tasks.

(1) The orthodoxy test should be applied to a wider range of published FN plots, in order to establish the presence of any systematic trends.

(2) For emitters that have tested as "effectively orthodox", improved methods need to be developed for extracting information about formal area efficiency $\alpha_f$, so that we can obtain better understanding of the range of values that this parameter might take.

(3) The effect of series resistance on measured CFE current-voltage characteristics needs to be explored *experimentally* in well-controlled situations (for example, a known resistance value in series with a STFE of moderate-to-large apex radius, whose characteristics are known), and compared with simulations.

(4) The procedure of "phenomenological adjustment" needs to be applied to a range of published FN plots found to be non-orthodox.

(5) The consequences of using semiconductor-like band-structures, rather than metal-like band-structures, in theoretical discussions of FN plot interpretation need to be explored.

More generally, this work confirms our view that FN-plot analysis is best done by using the raw experimental data and $i_m(V_m)$-form plots (or alternatively $J_M(V_m)$-form plots), with FEF-values deduced from extracted LCL values (where reliable values are available).

## Acknowledgement

This paper is based in part on material presented at the International Conference (Humboldt Kolleg) on "Building International Networks for Enhancement of Research in Jordan", held at Princess Sumaya University for Technology, Amman, Jordan, in April 2014.

## Appendix A: Estimation of the uncertainty limits for $\lambda_L$

For the new-standard FN-type equation, where tunnelling takes place through a SN barrier, estimates can be made of the range within which the pre-exponential correction factor $\lambda_L$ lies. As shown in Table 4, $\lambda_L$ is the product of components, most of which have a range of variability or uncertainty associated with them. Values for the first three rows derive from calculations by Mayer [36]; a formula for temperature effects was first given by Murphy and Good [12]; the uncertainty related to row 5 comes from information provided by Modinos [37,38]. This table is an improved version of one presented some years ago [39]. The values here can also be taken as "first estimates" applying to the value of $\lambda_L$ in the general FN-type equation. There is some reason to think that, in comparison with more general "phase-integral" [40] methods for evaluating transmission probabilities, the methods normally used in FE theory (described earlier) are first-order approximations; consequently, Table 4 may in fact underestimate the true range of uncertainty.

TABLE 4. Estimated values for $\lambda_L$ and its components (for "orthodox" FN-type equation based on SN barrier) (estimates made in September 2014).

| Row No. | Physical origin of correction-factor component | Symbol for component | Value of correction-factor component | Source |
|---|---|---|---|---|
| 1 | Tunnelling pre-factor | $P_F$ | ~ (0.4 to 1.1) | [36] |
| 2 | Correct summation over states | $\lambda_D$ | ~ (0.9 to 1.3) | [36] |
| 3 | Combination of above effects[a] | $\lambda_D P_F$ | ~ (0.5 to 1.0) | [36] |
| 4 | Temperature effects at 300 K | $\lambda_T$ (300 K) | ~ 1.1 | [12] |
| 5 | Electronic effects [atomic wave-functions & band-structure effects] | $\lambda_E$ | ~ (0.01 to 10) | [37,38] |
| 6 | All effects together | $\lambda_L$ | ~ (0.005 to 11) | |

[a]Note that high values of $\lambda_D$ tend to be associated with low values of $P_F$.

# Appendix B: Definition of the orthodox field electron emission measurement situation

A field electron emission measurement situation is termed "orthodox" if the following conditions are adequately satisfied.

(1) The "emission voltage" (i.e., the voltage between the emitting surface region and the counter-electrode) can be treated as uniform across the emitting surface and equal to the measured voltage.

(2) The emission current can be treated as equal to the measured current, with this measured current being controlled solely by the emission at the emitter surface.

(3) Emission can be treated as if it involves deep tunnelling through a Schottky-Nordheim barrier, with the emission current described by a related FN-type equation in which the independent variable is, or is exactly proportional to, the measured voltage, and the only equation parameter depending on the measured voltage is the barrier-form correction factor.

(4) The emitter local work function is constant across the emitting region, is constant in time, is current independent, and is equal to its assumed value.

These conditions are equivalent to assuming that, in the universal FN-type equation, the parameters $\phi$, $c_X$, $c_Y$ and $C_{YX}$ may be treated mathematically as constants, $\phi$ has the correct value, and $v_F^{GB}$ may be set equal to the SN-barrier-function value $v_F$.

# Appendix C: Definitions and approximations for subsidiary SN-barrier functions

All the SN-barrier functions can be obtained from the principal SN-barrier function $v(f)$, discussed in Ref. [15]. The definitions and approximated expressions [based on Eq. (12)] given here are adapted from Ref. [11]. For Eq. (B4), values of $\eta$ need to be derived from Eq. (13).

$$v(f) \approx 1 - f + (f/6)\ln f. \quad (B1)$$

$$u(f) = -dv/df \approx (5 - \ln f)/6. \quad (B2)$$

$$s(f) = v - f dv/df = v + uf \approx 1 - f/6. \quad (B3)$$

$$\left.\begin{array}{l} r_{2012}(f) = \exp[\eta u] \\ \approx \exp[(\eta/6)(5 - \ln f)] \end{array}\right\}. \quad (B4)$$